# A NOVEL METHODLOGY FOR THERMAL ANALYSIS & 3-DIMENSIONAL MEMORY INTEGRATION

Annmol Cherian[1], Ajay Augustine[1], Jemy Jose[2] and Vinod Pangracious[2]

[1]Department of Applied Electronics & Instrumentation, Rajagiri School of Engineering & Technology, Kerala, India.
annmolcherian26@gmail.com , ajayaugustine@hotmail.com

[2]Department of Electronics & Communication, Rajagiri School of Engineering & Technology, Kerala, India
jemyjosek@gmail.com , pangracious@googlemail.com

*Abstract*

*The semiconductor industry is reaching a fascinating confluence in several evolutionary trends that will likely lead to a number of revolutionary changes in the design, implementation, scaling, and the use of computer systems. However, recently Moore's law has come to a stand-still since device scaling beyond 65 nm is not practical. 2D integration has problems like memory latency, power dissipation, and large foot-print. 3D technology comes as a solution to the problems posed by 2D integration. The utilization of 3D is limited by the problem of temperature crisis. It is important to develop an accurate power profile extraction methodology to design 3D structure. In this paper, design of 3D integration of memory is considered and hence the static power dissipation of the memory cell is analysed in transistor level and is used to accurately model the inter-layer thermal effects for 3D memory stack. Subsequently, packaging of the chip is considered and modelled using an architecture level simulator. This modelling is intended to analyse the thermal effects of 3D memory, its reliability and lifetime of the chip, with greater accuracy.*

## 1.1  3-DIMENSIONAL OR VERTICAL INTEGRATION

Successful fabrication of vertically integrated device dates back to early 1980s. The structures include 3D CMOS inverters where PMOS and NMOS transistors share the same gate, considerably reducing total area of the inverter. The term Joint Metal Oxide Semiconductor (JMOS) was used for these structures. In the following years the result on 3D integration remained an area of limited scientific interest. Due to increasing importance of interconnects and the demand for greater functionality on a single substrate, vertical integration has recently become a more prominent research topic. Over the last 5 years, the 3D integration has evolved into a design paradigm manifested at several abstraction levels such as package, die and wafer levels. The quintessence of 3D integration is the drastic decrease in length of the longest interconnects across integrated circuits. The considerable decrease in length is a promising solution for increasing the speed while reducing the power dissipated by the IC.

## 1.2 CHALLENGES FOR 3D INTEGRATION

Developing a design flow for 3D ICs is a complicated task with many ramifications. Number of challenges at each step of the design process has to be satisfied for 3D ICs to successfully evolve into a mainstream technology. An efficient front-end design methodology and mature manufacturing preprocess at the back end are required to effectively provide large scale 3D systems. Challenges in this field are explained below.

- Technological/manufacturing limitations- new packaging solutions and utilization methods of vertical interconnects for the propagate signals are to be found out.
- Testing of 3D ICs which include die-to-die, wafer-to-wafer or die-to-wafer is more complicated than that of normal ICs.
- Interconnect design and new design methodologies are to be found out to combine different layers.
- Thermal issues - the power dissipation is increased due to increased transistor density. Due to increased power density, the part of IC away from heat sink has much elevated temperature and hence can accelerate wear out. These are the fundamental issues to be solved.

## 2.1 THERMAL MODELING & MANAGEMENT

The primary advantage of 3D integration, significantly the greater packing density, is also the greatest threat to this emerging technology - aggressive thermal gradient among the planes within a 3D IC. Thermal problems however are not unique to vertical integration. Due to scaling, elevated temperatures and hot spots within traditional 2D circuits, can greatly decrease the maximum achievable speed and affect the reliability of a circuit.

In 3-dimensional integration, low operating temperature is a prominent design objective, as thermal analysis of 3D ICs indicates the escalated temperatures can be highly problematic. Two key elements are required to establish a successful thermal management strategy: the thermal model to characterize the thermal behavior of a circuit and design techniques that alleviate thermal gradients among the physical planes of a 3D stack while maintaining the operating temperature within the acceptable levels. The primary requirements of a thermal model are high accuracy and low computational time, while thermal design techniques should produce high quality circuits without incurring long computational design time.

## 2.2 THERMAL ANALYSIS OF 3D ICs

A 3D system consists of disparate materials with considerably different thermal properties, including semiconductor, metal, dielectric and possibly polymer layers used for plane bolding. To describe the heat transfer process (only by conduction) within the volume of a system and determine the temperature at each point, T at a steady state requires a solution of the equation,

$$\nabla(k\nabla T) = -Q \tag{1}$$

Where k is the thermal conductivity and Q is the heat generated. In integrated circuits, heat originates from the transistor that behaves as a heat source and also from self heating of both the devices and the interconnects (joule heating), which can significantly elevate the circuit temperatures. More specific technique applied at various stages of IC design flow, such as synthesis, floor planning and placement, routing and maintaining the temperature of a circuit within the specified limits of or elevates thermal gradients among the planes of the 3D circuits. Many solutions are available for equation (1) which requires unacceptable computational time. To alleviate these issues standard methods to analyze heat transfer is used. Simple analytic expressions are developed. They are explained in the next section.

### 2.2.1 Closed form Temperature Expressions

A 3D system can be modeled as a cube consisting of multiple layers of silicon, aluminum, silicon dioxide and polyimide. The device on each plane is considered as isotropic heat source and each is modeled as a thin layer on the top surface of the silicon layer. In addition, due to the short height of 3D stack, one dimensional heat flow is assumed. Such an assumption vastly decreases the simulation time. Certain boundary conditions apply to this thermal model in order to validate the assumption of one dimensional heat flow. Self-heating of MOSFET devices can also cause the temperature of a circuit to rise significantly. Certain devices can behave as hot spots, causing significant local heating. Although the dielectric and glue layers behave as thermal barriers, the silicon substrate of the upper planes spreads the heat, reducing the self heating of the MOSFETs.

To estimate the maximum rise in temperature on the upper planes of a 3D circuit while considering the heat removal properties of interconnects, a simple closed form expression based on one dimensional heat flow has been developed. The temperature increase $\Delta T$ in a 2D circuit can be described by,

$$\Delta T = R_{th} P/A, \qquad (2)$$

where $R_{th}$ is the thermal resistance between the ambient environment and the actual devices, P is the power consumption and A is the circuit area.

### 2.2.2 Compact thermal models

Thermal models based on analytic expressions for the temperature of a 3D circuit are discussed. In all of these models, the heat generated within each physical plane is represented by a single value. Consequently, the power density of a 3D circuit is assumed to be a vector in the vertical direction (i.e.: z direction). The temperature and heat within each plane of 3D system however can vary considerably, yielding temperature and power density vectors which depend on all three directions. The nodes are connected through resistors forming a 3D resistive network.

### 2.2.3 Mesh based thermal models

This type of model most accurately represents the thermal profile of a 3D system. As a primary advantage of mesh-based thermal models, these models can be applied to any complex geometry and do not depend on the boundary conditions of the problem. A 3D circuit is decomposed into a 3D structure of finite hexahedral elements. (i.e.: parallelepipeds). The mesh can also be non-uniform for regions with complex geometries or non-uniform power densities.

## 3.1 CHIP POWER EXTRACTION & ESTIMATION

The MOS technologies have been improving at a dynamic rate. As the continuation of Moore's law, the MOS devices are scaled down aggressively in each generation to achieve higher integration density and performance. Thus, the recent trend in semiconductor manufacturing is to build devices in nanoscale. The major problem with nanoscale devices is the increasing power dissipation. The total static power of CMOS technology is comprised of two major components: the device subthreshold current $I_{OFF}$ and the gate tunneling current $I_{GATE}$. Therefore, an accurate analysis of the total static power must include both the channel leakage due to device subthreshold current and the gate leakage due to tunneling current. A brief introduction is given to each of these effects which arise when scaling down the MOS devices.

### 3.1.1 Sub-threshold leakage

From its earlier days, MOS devices were well known for the switching applications. By applying suitable gate voltage, the current flow from the source to drain is controlled. Moreover, by the application of gate voltage less than the threshold voltage, the device would be turned off and resulting in zero drain current.

The problem arises when the whole semiconductor industry is shrinking to nanometerscale. When the devices were fabricated in nanometer scale, the electric field increased at constant rate, as the voltage drops over the gate oxide and the channel stayed the same when their sizes were reduced, leading to reliability concerns. The supply voltage was reduced to overcome these problems. Consequently, the threshold voltage was reduced. As a result, the off state current increased and gradually became a limiting factor in scaling of MOS devices. Therefore, while modeling a nanometer scale device the "subthreshold" behavior has to be considered.

Basically there are three different regions of MOS operation based on the inversion. They are weak inversion, moderate inversion and strong inversion. The current flow can be due to two reasons - drift and diffusion. Under weak inversion the channel surface potential is almost constant across the channel and the current flow is determined by diffusion of minority carriers due to a lateral concentration gradient. Under strong inversion there exists a thin layer of minority carriers at the channel surface and a lateral electric field, which causes a drift current. The moderate inversion regime is considered as a transition region between weak and strong inversion where both current flow mechanisms coincidently exist. In the weak-inversion (or subthreshold) region, the drain current depends exponentially on the gate-source voltage. The exponential subthreshold behavior is due to the exponential dependence of the minority carrier density on the surface potential, which itself is proportional to the gate voltage. On a semi-logarithmic scale the transfer (or $I_d$-$V_g$) curve in the subthreshold region will, therefore, be a straight line.

### 3.1.2 Gate leakage

As the device is scaled down, the gate oxide thickness is also reduced. Therefore, the leakage current can flow from channel to gate attaining prominence. Thus the gate cannot be considered as perfectly insulated. This affects the circuit functionality and increases the static power. Also the gate leakage reduces the clock cycle time. The two tunneling mechanisms which cause gate leakage are Fowler-Nordheim tunneling and direct tunneling. Direct tunneling means tunneling in nano - MOS structures in which electrons from the conduction band in semiconductor are transferred across the oxide directly (i.e. without changing energy) into the conduction band of metal; probability of direct tunneling is a very strong function of the width of the barrier through which electron tunnels (oxide thickness in MOS devices). Fowler-Nordheim means tunneling of electrons from semiconductor conduction band into oxide conduction band, through a part of the potential barrier at the semiconductor-oxide interface; most likely to dominate in MOS structures with oxide 5-10 nm thick. The gate leakage increases exponentially as the oxide thickness is reduced. This limits the down-scaling of the oxide thickness to about 1.5-2nm when looking at the total standby power consumption of a chip.

### 3.1.3 Hot carrier effects

When a MOS transistor is operated in pinch-off condition, also known as 'saturated case', hot carriers traveling with saturation velocity can cause parasitic effects at the drain side of the channel. This is known as 'Hot Carrier Effects' (HCE). These carriers have sufficient energy to generate electron-hole pairs by Impact ionization. The generated bulk minority carriers can either be collected by the drain or injected into the gate oxide. The generated majority carriers create a bulk current which can be used as a measurable quantity to determine the level of impact ionization. Carrier injection into the gate oxide can also lead to hot carrier degradation effect. This happens when the threshold voltage changes due to the occupied traps in the oxide. Hot carriers can also generate traps at the silicon-oxide interface known as 'fast surface states', leading to sub-threshold swing deterioration and stress-induced drain leakage. In general, these degradation effects set a limit to the lifetime of a transistor, leading to the necessity of their control.

### 3.1.4 Drain Induced Barrier Leakage

In long channel devices, the gate is completely responsible for depleting the semiconductor. But in very short channel devices, part of the depletion is accomplished by the drain and source bias. This is shown in figure (1). Since less gate voltage is required to deplete the channel, threshold voltage decreases as length decreases. Similarly, as the drain voltage is increased, more substrate is depleted by the drain bias, and hence threshold voltage decreases. These effects are particularly pronounced in lightly doped substrates.

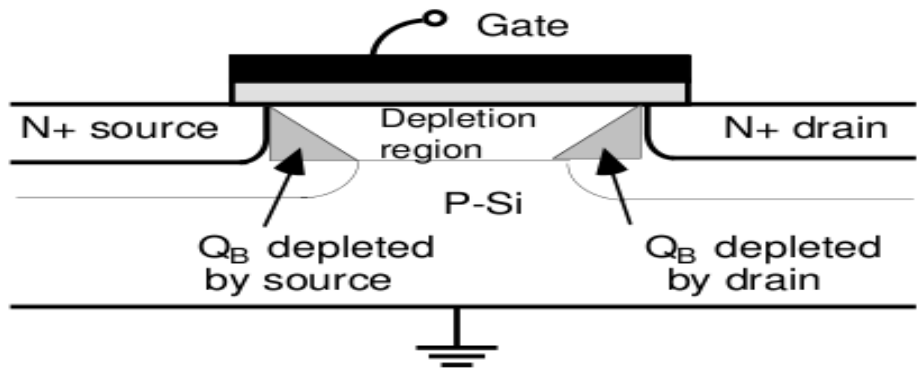

Figure 1. Normal operation of PMOS

If the channel length becomes too short, the depletion region from the drain can reach the source side and reduces the barrier for electron injection. This is known as punch through.

In devices with long channel lengths, the gate is completely responsible for depleting the semiconductor. In very short channel devices, part of the depletion is accomplished by the drain and source bias. Since less gate voltage is required to deplete semiconductor, the barrier for electron injection from source to drain decreases. This is known as Drain Induced Barrier Lowering (DIBL). DIBL results in an increase in drain current at a given gate voltage. Therefore threshold voltage decreases as length decreases. Similarly, as drain voltage increases, more semiconductors are depleted by the drain bias, and hence drain current increases and threshold voltage decreases.

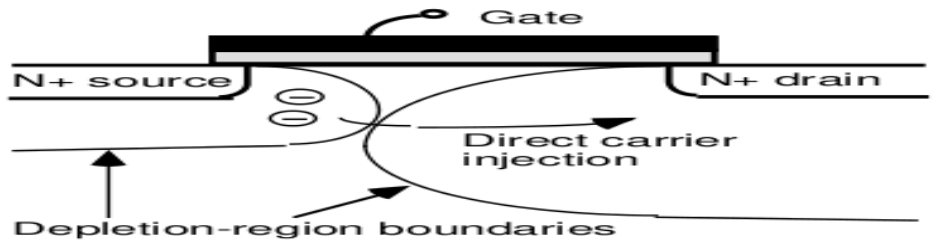

Figure 2. DIBL effect

### 3.2. BUTTS AND SOHI MODEL

The first experimental result regarding the sub-threshold leakage was the Butts and Sohi model. This presents a high-level model of sub-threshold leakage that neatly compartmentalizes some different issues affecting static power in a way that makes it easy to reason about leakage effects at the micro-architecture level. But it doesn't include the leakage effects due to temperature. Even then, the Butts and Sohi model provides a starting point for the study. The power dissipation according to Butts and Sohi is given by,

$$P_{static} = V_{cc} \times N \times K_{design} \times I_{leakage} \tag{3}$$

$V_{CC}$ is the supply voltage, and N is the number of transistors in the circuit, which could be estimated by comparing it with a circuit of known functionality. $K_{design}$ is a factor determined by the specific circuit topology and accounts for effects like transistor sizing, transistor stacking and the number and relationship of NMOS and PMOS transistors in a circuit. $I_{leakage}$ is a normalized leakage value for a single transistor - can be named as a unit leakage and includes all technology-specific effects like threshold voltage ($V_T$) and also factors in the operating temperature. This model helps in finding various parameters which can be used to control the effect of leakage power savings since $I_{leakage}$ is proportional to the operating voltage and number of transistors used. Unfortunately this will have only little usage in today's manufacturing process. The effect of $V_{dd}$ and temperature are not included in this model. and so it is not suited for actual simulations.

## 3.3. BSIM MODEL

In order to get an accurate model that includes the different leakage parameters, a detailed literature survey was done. The leakage model of a single transistor was found as given below:

$$I_{leakage} = \mu 0 \times C_{ox} \times (w + l) \times e^{b(vdd-vdd0)} \times vt^2 \times (1 - e^{-vdd \div vt})$$
$$\times e^{(-|vth|-Voff) \div n \times vt} \quad (4)$$

Low-level parameters are derived using transistor-level simulations: $\mu 0$ is the zero bias mobility, $C_{OX}$ is gate oxide capacitance per unit area, W/L is the aspect ratio of the transistor, $e^{b(vdd-vdd0)}$ is the DIBL factor derived from the curve fitting method, $V_{dd0}$ is the default supply voltage for each technology ($V_{dd0} = 2.0$ for 180nm, $V_{dd0} = 1.5$ for 130nm, $V_{dd0} = 1.2$ for 100nm and $V_{dd0} = 1.0$ for 70nm), $v_t = kT/q$ is the thermal voltage, $V_{th}$ is threshold voltage which is also a function of temperature, n is the subthreshold swing coefficient, $V_{off}$ is an empirically determined BSIM3 parameter which is also a function of threshold voltage. In these parameters, $\mu 0$, $C_{OX}$, W/L and $V_{dd0}$ are statically defined parameters; the DIBL factor b, subthreshold swing coefficient n and $V_{off}$ are derived from the curve fitting method based on the transistor level simulations; $V_{dd}$, $V_{th}$ and $vt - kT \div q$ are calculated dynamically.

The above equation is based on two assumptions:

1. $V_{gs} = 0$ — considering only the leakage current when the transistor is off.
2. $V_{ds} = V_{dd}$ — considering only a single transistor here; the stack effect and the interaction among multiple transistors are taken into account when the cell is modeled using Equation (3)
The $V_{th}$ in the above equation was extracted from BSIM3 paper and is given by,

$$V_{th} = \underbrace{V_{th0ox}}_{A} + \underbrace{K_{1ox} \cdot \sqrt{\Phi_s - V_{bseff}}}_{B} - \underbrace{K_{2ox} V_{bseff}}_{C}$$
$$+ \underbrace{K_{1ox}\left(\sqrt{1 + \frac{Nlx}{L_{eff}}} - 1\right)\sqrt{\Phi_s}}_{D} + \underbrace{(K_3 + K_{3b}V_{bseff})\frac{T_{ox}}{W_{eff}' + W_0}\Phi_s}_{E}$$
$$- \underbrace{D_{VT0w}\left[\exp\left(-D_{VT1w}\frac{W_{eff}'L_{eff}}{2l_{tw}}\right) + 2\exp\left(-D_{VT1w}\frac{W_{eff}'L_{eff}}{l_{tw}}\right)\right](V_{bi} - \Phi_s)}$$
$$- \underbrace{D_{VT0}\left[\exp\left(-D_{VT1}\frac{L_{eff}}{2l_t}\right) + 2\exp\left(-D_{VT1}\frac{L_{eff}}{l_t}\right)\right]}_{F}(V_{bi} - \Phi_s)$$
$$- \underbrace{\left[\exp\left(-D_{sub}\frac{L_{eff}}{2l_{to}}\right) + 2\exp\left(-D_{sub}\frac{L_{eff}}{l_{to}}\right)\right]}_{H}\underbrace{(E_{tao} + E_{tab}V_{bseff})V_{ds}}_{G}$$

Figure 3.

The parameters mentioned are
$V_{th0}$ : Threshold voltage at $V_{sb} = 0$
$K_1$ : 1st order body effect coefficient

| | |
|---|---|
| $K_2$ | : $2^{nd}$ order body effect coefficient |
| $K_3$ | : NW coefficient |
| $\phi_s$ | : Surface potential |
| Nlx | : Lateral non-uniform doping parameter |
| $K_{3b}$ | : Body effect coefficient of $K_3$ |
| $W_0$ | : NW parameter |
| $D_{vt0w}$ | : $1^{st}$ coefficient of NW effect on $V_{th}$ for SCL |
| $D_{vt1w}$ | : $2^{nd}$ coefficient of NW effect on $V_{th}$ for SCL |
| $V_{bi}$ | : Built-in voltage of PN junction between S-S |
| $D_{vt0}$ | : $1^{st}$ coefficient of SCL effect on $V_{th}$ |
| $D_{vt1}$ | : $2^{nd}$ coefficient of SCL effect on $V_{th}$ |
| $D_{sub}$ | : DIBL coefficient exponent in ST region |
| $E_{ta0}$ | : DIBL coefficient in ST region |
| $E_{tab}$ | : Body bias coefficient for ST DIBL effect |

For a specific cell, the leakage current is given by the following equation:

$$I_{cellleakage} = (n_N \times I_N + n_P \times I_P) K design \qquad (5)$$

$n_N$ and $n_P$ are the number of NMOS and PMOS transistors in the cell, and $I_N$ and $I_P$ are the calculated leakage current of NMOS and PMOS; when aspect ratio W/L = 1, it is unit leakage; $k_{design}$ is the design factor determined by the stack effect and aspect ratio of transistors. $K_{design}$ is derived from transistor-level simulation of an actual design and layout of a cell of interest. Given a cell, the average leakage $I_{cellleakage}$ is derived from the transistor-level simulation with all possible input combinations. $K_{design}$ is the factor which accounts for the transistor aspect ratio (W/L) and the stack effect. (The stack effect refers to the additional reduction in leakage when multiple series-connected transistors are off; for example, sleep transistors take the advantage of this.) Unlike the Butts and Sohi model, here the $K_{design}$ does in fact vary with temperature, supply voltage, threshold voltage and channel length.

### 3.4. IMPROVED MODEL

The single $K_{design}$ model is suitable for cases where the parameters of N and P transistors are very close. If these two sets parameters of N and P transistors differ significantly, different $K_{design}$ should be applied to these two types of transistors. Thus the improved leakage model: double-$K_{design}$ model is used. For a specific cell, the leakage current is now given by this different equation:

$$I_{cellleakage} = n_n \times k_n \times I_n + n_p \times k_p \times I_p \qquad (6)$$

$K_n$ and $k_p$ are the design factors of N and P transistors and they can be derived by the same method as in the single- $K_{design}$ model. For a given cell, divide all possible inputs into two groups: one group inputs will turn off the pull-down network composed of N transistors. The other group will turn off the pull-up network composed of P transistors. Thus the leakage currents are also divided into two groups $I_{1n}, I_{2n}, \ldots, I_{kn}, \ldots$ and $I_{1p}, I_{2p}, \ldots, I_{kp}, \ldots$ where $I_{kn}$ is the leakage current when the pull-down network is turned off, while $I_{kp}$ is the leakage current when the pull-up network is turned off. $k_n$ and $k_p$ are given by the following equations.:

$$K_n = (I_{1n} + I_{2n} + \cdots + I_{kn} + \cdots) \div (N \times n_n \times I_n) \qquad (7)$$
$$k_p = (I_{1p} + I_{2p} + \cdots + I_{kn} + \cdots) \div (N \times n_p \times I_p) \qquad (8)$$

N is the number of all possible combinations. The improved leakage model is used to estimate the leakage current of a single transistor.

## 4.1. 3D MEMORY SIMULATION & MODELING

To model the 3D memory accurately, modeling is started with the basic circuitry. The cache memory consists of SRAM cells. The SRAM has a 6 transistor structure. The leakage current of this arrangement is calculated using the double $K_{design}$ methodology, explained in the initial part of the paper. This leakage current is then used for the 2D and 3D analysis. The leakage current obtained by this method includes all the effects which were explained earlier. So, this is the most accurate method for finding leakage current.

The software which is used for simulation is Hotspot and STiMuL. Hotspot is an embedded tool that takes power dissipation and floor plan as the inputs and gives the temperature profile of the arrangements. The power dissipation given is based on the static power extraction methodology described earlier. Both two dimensional and three dimensional circuits can be modeled using this. The results are tabulated. From this, the memory cells are modeled and the temperature profiles are obtained. The temperature profile obtained in this method is more accurate because the process is started from the circuit level.

Now the thermal model while packaging is to be considered. For this the STiMuL software is used. STiMuL provides the ultimate model for packaging. It includes modeling of 2D structures and also packaging. According to the input parameters the output plot is obtained which gives the temperature distribution along the distance. In this paper first the model without packaging effect is studied. The results are compared with those obtained in Hotspot. Then the packaging is included and the results are analyzed. Detailed explanation is given in the next section.

## 5.1. EVALUATION AND RESULTS

The evaluation process is divided into three.
- The first part consist of modeling of a single transistor considering all the leakage effects such as sub-threshold leakage, gate leakage, hot carrier effects and DIBL. The results include calculation of leakage current of PMOS, NMOS, CMOS and SRAM cell. The plots for variation in leakage current with $V_{dd}$ and variation of leakage current with temperature are obtained. This makes the first stage of the evaluation and is named as thermal modeling.
- In the second step the thermal profile of conventional 2D memory circuits is obtained using Hotspot. Here each memory cell is having a power dissipation of 10W. For 3D circuits seven cases are considered and the results are explained below in detail. This step is named as circuit modeling.
- As the third step the whole packaging is to be modeled. For this STiMuL is used. Initially 2D circuits are modeled to get a familiarization of the software. In STiMuL the heat generating areas are given as the heat sources and the corresponding plots are obtained. This step is named as package modeling.

## 5.2. THERMAL MODELING

The leakage current of transistor level circuit was found out based on the conclusions from literature survey. Based on the leakage model a code was written in C language. The aim of the code was to obtain the leakage current. An option was introduced to select among NMOS and PMOS. The parameters such as $V_{dd}$ and temperature can be changed. Thus the plots of leakage current with variation in temperature and $V_{dd}$ are plotted.

For the 3D simulation the leakage current of an SRAM cell is also calculated. The 6T structure of SRAM cell is considered. The results are given below.

Leakage current of NMOS transistor ($I_{leakage}$) =8.585 x $10^{-7}$ A

Leakage current of PMOS transistor ($I_{leakage}$)     =1.692 x $10^{-7}$ A
Leakage current of SRAM cell ($I_{cellleakage}$)     =1.027 x $10^{-6}$ A

The variation of leakage current with temperature and supply voltage is found in order to check the accuracy of the thermal modelling process. If the curve follows the ideal curve then the strategies taken are perfect. In this case both the characteristics follow the ideal curve and so the strategies are accurate and perfect. Leakage current Vs temperature graph is given below.

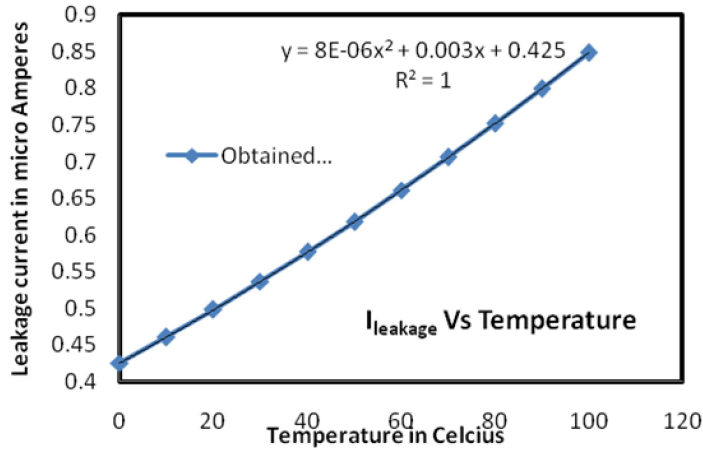
Figure 4. Temperature Vs leakage current graph

The above plot is the variation of leakage current with temperature. The ideal behavior of leakage current with temperature is a super linear curve. The result obtained from our simulation follows the ideal curve. Temperature is varied from 0°C to 100°C by keeping $V_{dd}$ constant and $I_{leakage}$ is calculated. The figure 5 shows the plot for leakage current Vs supply voltage.

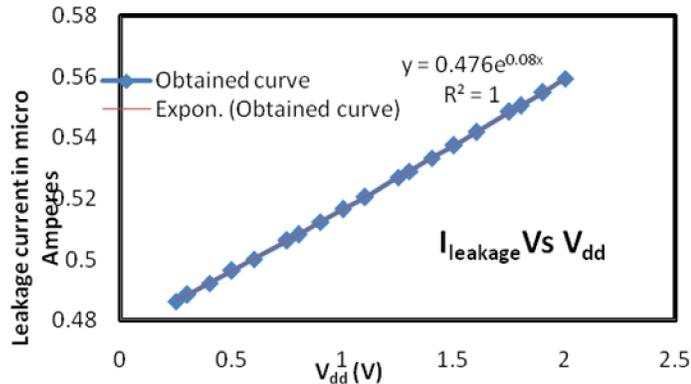
Figure 5. Supply voltage Vs leakage current graph

The above plot is the variation of leakage current with $V_{dd}$. Supply voltage is varied from 0 V to 2 V keeping temperature constant and leakage current was calculated. The ideal characteristics should have a parabolic shape. The curve obtained from the simulation can be approximated to parabolic. Thus the simulation has circuit level accuracy.

## 5.3. CIRCUIT MODELING

The circuit modeling process consists of finding the temperature of all inner circuit elements. This step was accomplished by the software HOTSPOT. Both 2D and 3D simulation was possible in Hotspot. The 2D simulation results were plots and 3D results were temperature values.

### 5.3.1. 2D Modeling

Initially 2D circuit models are used for simulation. The circuit has four cache memory, eight other circuit components which can be application specific depending on the manufacturer and a crossbar for interconnection. The power dissipation of other circuit is assumed to be uniformly low compared to the memory cells. Here the maximum power dissipation is given to the memory cells. Then the operating temperature of memory cell is found to be 331.26K. This range is permissible for proper operation of memory cell. The results and plots are shown below. The 3D simulation in next section is given more importance in this paper.

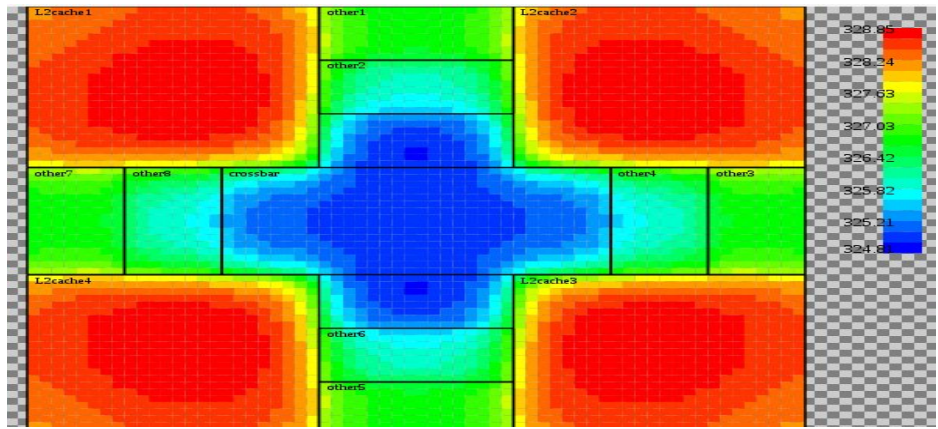

Figure 6. 2D simulation with 8 w power dissipation

### 5.3.2. 3D Modeling

The 3D modeling facility provided in the hotspot software gives the temperature in each layer. The highest temperature in each layer is taken for the comparative study. Five experiments are done for 3D memory model with six layer architecture. The six layer architecture means six memory layers are stacked vertically. The stack consists of six layers of silicon active layers, each separated by an epoxy layer. The epoxy layer is used mainly to isolate the silicon layer. Therefore totally there will be 11 layers. The floor plan for the 3D simulation is same as that of the 2D circuit. The same 2D circuits are stacked together to form 3D structure. An example of five layer architecture is given below.

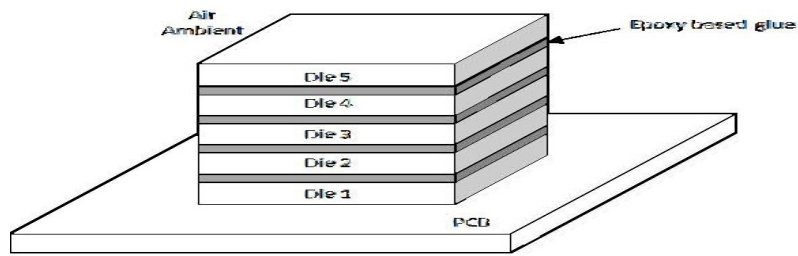

Figure 7. An example of 3D structure

The temperature with the change in power dissipation is to be studied. For this the power dissipation for memory cells can be changed and the resulting operating temperature is analyzed. Experiments 1, 2 & 3 are done for this. The structure considered is with 6 layers of silicon and with uniform power distribution for all the memory cells. The floor plan and power profile files are the same as that of the 2D. Here all the layers and all cells in the layers are powered. The power dissipation of memory cells are fixed to $3W/cm^2$ in experiment 1, $8W/cm^2$ in experiment 2 and $10W/cm^2$ for experiment 3. The resistivity of the epoxy is kept constant. The memory cells are given the highest power dissipation and all other circuits are given a lesser value. The above three experiments ranges from the worse case to the normal working range of power dissipation. With the change in power dissipation it is found to be increasing. Even though there is a change, the decrease in temperature is uniform in all the three cases. The experiment 2 is only considered and plotted in the bar chart. The temperature level is decreased as the number of layers is increased. This is because of the lateral spreading of the heat generated in the cells and no boundary conditions are specified.

The next aim is to reduce the working temperature. For this, the dynamic thermal strategy is used. When the 3D structure is working it is not necessary that all the layers work simultaneously. There are some strategies to reduce the operating temperature by selective powering of the memory cells. One such method is diagonal powering of memory cell. Other methods are selective placing of data, selective reading etc. To accomplish this, in experiment 4, all the memory cells are not powered simultaneously, but the alternate layers are powered. When the alternate layers are powered the temperature value decreased and the characteristic decrease in temperature is different from all other experiments. In experiment 5, another strategy used is that only one layer is powered. The temperature value further decreased in this case and here also the response is different. Even though all the layers are not working, the heat generated in one layer is transferred to another. That is why the upper layer has an increased temperature. Another study is done to know what happens when only two cells in a layer are accessed. The response is more or less similar to the previous experiment since both have only simple changes.

Now the behavior of the epoxy layer has to be studied. The epoxy layer has been inserted to provide a glue interfacing. So, the characteristics has to be determined if the epoxy layer is changed. The epoxy resistivity is changed in order to study the effect of heat transfer with different materials used as glues. This also helps in transferring the hotspot temperature created during the operation to another spot so that the decay of cell due to hotspot creation is avoided. Three experiments are done to study the effect TSV (through silicon vias) interconnects for 3D stacked chip. When the epoxy materials change, it has effect on operating temperatures. The temperature value changes when the epoxy resistivity changes from 0.25 to 0.005. But not much change is seen when the epoxy values change from 0.005 to 0.001. Also the characteristic behavior is changed.

The results are tabulated and are given below. The plots for all the five experiments are given below.
Experiment 1: 3D- 6 layer-uniform-$3W/cm^2$- epoxy resistivity=0.25-all powered
Experiment 2: 3D- 6 layer-uniform-$8W/cm^2$-epoxy resistivity=0.25-all powered
Experiment 3: 3D- 6 layer-uniform-$10W/cm^2$-epoxy resistivity=0.25-all powered
Experiment 4: 3D- 6 layer-uniform-$3W/cm^2$-epoxy resistivity=0.25-alternate layers powered
Experiment 5: 3D- 6 layer-uniform-$3W/cm^2$-epoxy resistivity=0.25-layer1 powered
Experiment 6: 3D- 6 layer-uniform-$3W/cm^2$-epoxy resistivity=0.25-layer1 l2l4 powered
Experiment 7: 3D- 6layer-uniform-$3W/cm^2$-epoxy resistivity=0.001-all powered
Experiment 8: 3D- 6layer-uniform-$3W/cm^2$-epoxy resistivity=0.005-all powered

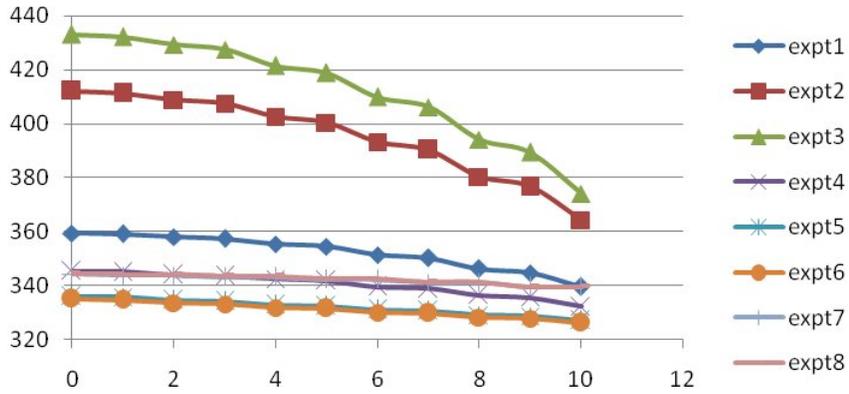

Figure 8. Results of all the experiments

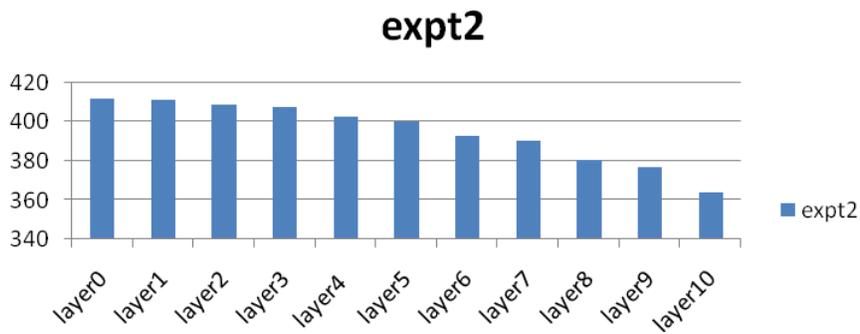

Figure 9. Experiment 2

### 5.3.3. Comparison with 3 layer silicon & 6 layer silicon

To study the effect of number of layers, a comparison is done between 3 layer architecture and 6 layer architecture. The results are shown in the table below. All the parameters except the number of layers are kept constant. As the number of layers increases the temperature at the bottom layer of the architecture rises sharply to higher degrees but, the decrease of temperature due to heat dissipation is more or less same with any number of layers.

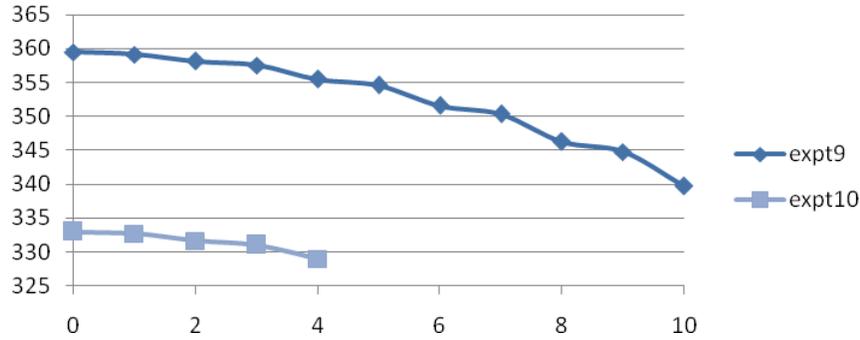

Figure 10. Comparison of 3 layers and 6 layers

Experiment 1: 3D- 3 layer-uniform-3w/cm2- epoxy resistivity=0.25-all powered
Experiment 1: 3D- 6 layer-uniform-3w/cm2- epoxy resistivity=0.25-all powered

## 5.4 PACKAGE MODELING

The circuit modeling with the hot spot software is compared with the results of StiMuL. Exact modeling of 3 D structures is not available with StiMuL and so only approximate modeling is done. The result obtained while simulating in StiMul is given below.

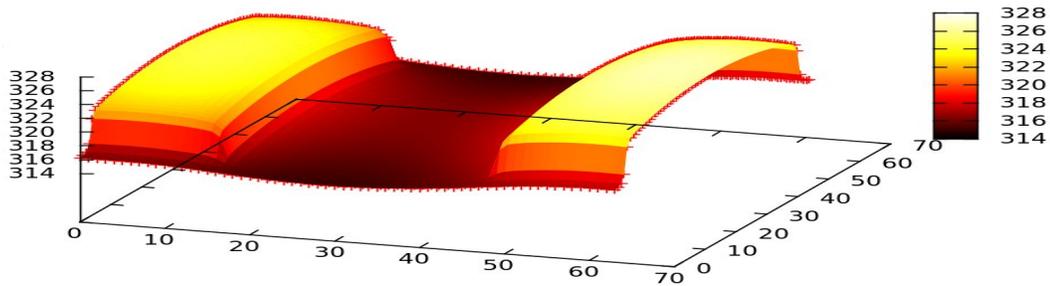

Figure 11. STiMul model o 3D memory cell

The next step is to include packaging and heat sink. The inclusion of heat sink should reduce the temperature dissipation. Also the packaging has to be considered to ensure the proper working of the chip. The simulation result obtained while including packaging is given below. The maximum temperature obtained in hotspot simulation is 331K. When the same condition is applied in StiMul the temperature becomes 328K. This is clear from the figure 13. When the heat sink and packaging is included the temperature is reduced to 318K. The obtained results are desirable since they lie in the required range.

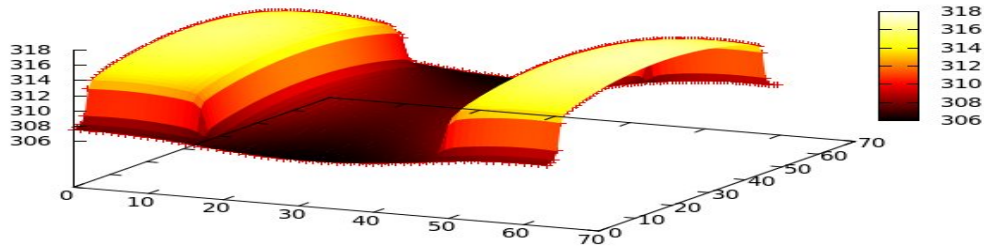

Figure 12. Packaging model of 3D memory cell

## CONCLUSIONS

This paper proposes the use of device and circuit level temperature dependent static leakage extraction methodology as an efficient and accurate methodology for modeling temperature induced thermal effects in 3D stacked memory chips. A thorough study of different types of leakage mechanisms in MOS transistor is conducted to formulate the leakage extraction methodology and the module is very much helpful to extract a fast and accurate power profile of a circuit. This paper discusses the detailed mechanism of how the temperature induced leakage extraction module can be interfaced with a standard 3D thermal modeling tool like Hotspot. By applying this methodology the most accurate thermal model for 2D and 3D integrated chips is obtained. Addressing of the complete product development issues of 3D integrated chip is tried through this paper by introducing the architectural level modeling and also package level modeling. This is an effective methodology for 3D integration, especially in circuits like memory. This methodology can also be extended to other areas like multi-processor system on chip (MPSoC).

## ACKNOWLEDGEMENTS

We would like to express our heartfelt gratitude to Dr. J. Isaac, Principal, Ms. Liza Annie Joseph, Head of the Department, Applied Electronics & Instrumentation, and Prof. Asha Paniker, Head of the Department, Electronics & Communication Engineering, Rajagiri School of Engineering & Technology, Kerala, India.

## REFERENCES

[1] Vasilis.F.Pavlidis, Eby.G. Friedman,"3d integrated circuit design", Morgan Kaufmann series in system on silicon.
[2] John F. McDonald,"Thermal and Stress Analysis Modeling for 3D Memory over Processor Stacks" ,Rensselaer Polytechnic Institute, Troy, NY 12180, "SEMATECH Workshop on Manufacturing and Reliability Challenges for 3D IC's using TSV's".
[3] Stephen Tarzia, "A Survey of 3D Circuit Integration", March 14, 2008.
[4] Igor Loi, Luca Benini," An efficient distributed memory interface for many core platform with 3D stacked DRAM", University of bologna,Italy.
[5] H.Mangalam and K.Gunavathi "Gate and subthreshold leakage reduced SRAM cells"DSP Journal, Volume 6, Issue 1, September, 2006.
[6] Howard Chen, Scott Neely, Jinjun Xiong, Vladimir Zolotov, and Chandu Visweswariah "Statistical Modeling and Analysis of Static Leakage and Dynamic Switching Power" IBM Research Division, Thomas J. Watson Research Center.
[7] Yan Zhang, Dharmesh Parikh, Karthik Sankaranarayanan, Kevin Skadron, Mircea Stan "HotLeakage: A Temperature-AwareModel of Subthreshold and Gate Leakage for Architects" University of Virginia Charlottesville, March 2003.
[8] Weidong Liu, Xiaodong Jin, James Chen, Min-Chie Jeng, Zhihong Liu, Yuhua Cheng, Kai Chen, ansun Chan, Kelvin Hui, Jianhui Huang, Robert Tu, Ping K. Ko and Chenming Hu" BSIM3v3.2.2 MOSFET Model", University of California, Berkeley, 1999.

[9] Howard Chen, Scott Neely, Jinjun Xiong, Vladimir Zolotov, and Chandu Visweswariah" Statistical Modeling and Analysis of Static Leakage and Dynamic Switching Power",IBM Research Division.

[10] Pierre Michaud," STiMuL: a Software for Modeling Steady-State Temperature in Multilayers - Description and user manual", INSTITUT NATIONAL DE RECHERCHE EN INFORMATIQUE ET EN AUTOMATIQUE.

[11] S. Das, A. Chandrakasan, and R. Reif, "Design tools for 3-D integrated circuits," in Proc. of ASP-DAC, 2003.

[12] HOTSPOT, University of virginia, http://lava.cs.virginia.edu/HotSpot/HotSpot-HOWTO.htm.

**Authors**

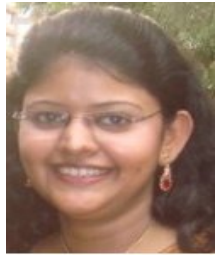

Annmol Cherian Vazhappilly completed B.Tech. degree in Applied Electronics & Instrumentation at Rajagiri School of Engineering & technology affiliated to Mahatma Gandhi University. The project currently working on is Thermal modeling & 3D integration of memory. She is an author of the paper named "Thermal Aware Static Power Extraction Methodology for Nanoscale Integrated Circuits". She has participated in national level conferences and presented the paper.

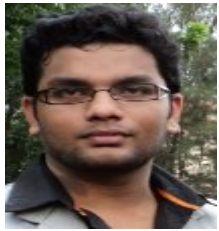

Ajay Augustine completed B.Tech. degree in Applied Electronics & Instrumentation at Rajagiri School of Engineering & technology affiliated to Mahatma Gandhi University. The project work undertaken was thermal modeling & 3D integration of memory. He is an author of the paper named "Thermal Aware Static Power Extraction Methodology for Nanoscale Integrated Circuits". Participated in national level conferences and presented the paper.

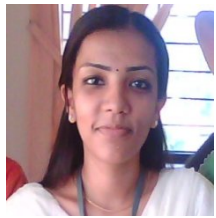

Jemy Jose Kakkassery completed B.Tech. in Electronics and Communication Engineering, from Amal Jyothi College of Engineering, affiliated to Mahatma Gandhi University, Kerala. She worked as a lecturer at Jyothi Engineering College for one and a half years. She is presently pursuing M.Tech. in VLSI and Embedded System, at Rajagiri School of Engineering and Technology, India. Project under taken was Automation of Partial Reflection Radar System using LabVIEW- from **VSSC-ISRO.** Currently doing research on Predictive Leakage Analysis on Nano-scale CMOS Technology.

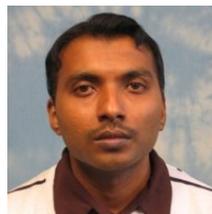

Vinod Pangracious received the B.Tech. degree in Electronics Engineering from Cochin University of Kerala in 1995 and M.Tech degree from IIT Bombay India in 2000. Currently pursuing Ph.D. at the University of Pierre and Marie Curie, Paris, France. He is currently an Associate Professor at Rajagiri School of Engineering & Technology India. He has authored and co-authored 10 publications in these areas. He is an internationally experienced Electronics Engineering professional with extensive expertise in memory design, high speed digital circuit design, logic library development, verification, test and characterization. His research interest focuses on the design methodologies for integrated systems, including thermal management technique for multiprocessor System on Chip, novel nanoscale architectures for logic and memories, 3D integration and manufacturing technologies.